\documentclass[12pt,nofootinbib]{article}

\usepackage{graphicx}
\usepackage{amssymb}
\usepackage{amsmath}
\usepackage{color}

\usepackage{graphicx}
\usepackage{bm}% bold math
\usepackage{hyperref}
\usepackage{color}

\usepackage{amsmath,amsfonts,amssymb}
\begin{document}

%%%%%%%%%%%%%%%%%%%%%%%%%%%%%%%%%%%%%%%%%%%

\def\o{\over}
\def\beq{\begin{align}}
\def\eeq{\end{align}}
\newcommand{\gsim}{ \mathop{}_{\textstyle \sim}^{\textstyle >} }
\newcommand{\lsim}{ \mathop{}_{\textstyle \sim}^{\textstyle <} }
\newcommand{\vev}[1]{ \left\langle {#1} \right\rangle }
\newcommand{\bra}[1]{ \langle {#1} | }
\newcommand{\ket}[1]{ | {#1} \rangle }
\newcommand{\EV}{ \text{eV} }
\newcommand{\KEV}{ \text{keV} }
\newcommand{\MEV}{ \text{MeV} }
\newcommand{\GEV}{ \text{GeV} }
\newcommand{\TEV}{ \text{TeV} }
\newcommand{\1}{\mbox{1}\hspace{-0.25em}\mbox{l}}
\newcommand{\headline}[1]{\noindent\textbf{#1}}
\def\diag{\mathop\text{diag}\nolimits}
\def\Spin{\mathop\text{Spin}}
\def\SO{\mathop\text{SO}}
\def\O{\mathop\text{O}}
\def\SU{\mathop\text{SU}}
\def\U{\mathop\text{U}}
\def\Sp{\mathop\text{Sp}}
\def\SL{\mathop\text{SL}}
\def\tr{\mathop\text{tr}}
\def\mpl{M_\text{Pl}}

\def\dd{\mathrm{d}}
\def\ff{\mathrm{f}}
\def\BH{\text{BH}}
\def\inf{\text{inf}}
\def\ev{\text{evap}}
\def\eq{\text{eq}}
\def\SM{\text{sm}}
\def\Mpl{M_\text{Pl}}
\def\GeV{\text{GeV}}
\newcommand{\Red}[1]{\textcolor{red}{#1}}

%%%%%%%%%%%%%%%%%%%%%%%%%%%%%%%%%%%%%%%%%%%%%%%%%%%%%%%%%%%%%%%
\begin{titlepage}
\begin{center}

\hfill IPMU-18-0027 \\
\hfill \today

\vspace{1.5cm}
\textbf{\Large
QCD axion dark matter from long-lived domain walls during matter domination
}

\vspace{2.0cm}
\textbf{Keisuke Harigaya}$^{(a,b)}$
and
\textbf{Masahiro Kawasaki}$^{(c,d)}$

\vspace{1.0cm}
\textit{%
$^{(a)}${Department of Physics, University of California, Berkeley, California 94720, USA}\\
$^{(b)}${Theoretical Physics Group, Lawrence Berkeley National Laboratory, Berkeley, California 94720, USA}\\
$^{(c)}${ICRR, University of Tokyo, Kashiwa, Chiba 277-8582, Japan}\\
$^{(d)}${Kavli IPMU (WPI), UTIAS, University of Tokyo, Kashiwa, Chiba 277-8583, Japan}
}

\vspace{2.0cm}
\abstract{
The domain wall problem of the Peccei-Quinn mechanism can be solved if the Peccei-Quinn symmetry is explicitly broken by a small amount. Domain walls decay into axions, which may account for dark matter of the universe. This scheme is however strongly constrained by overproduction of axions unless the phase of the explicit breaking term is tuned. We investigate the case where the universe is matter-dominated around the temperature of the MeV scale and domain walls decay during this matter dominated epoch. We show how the viable parameter space is expanded.
}
\end{center}
\end{titlepage}
\setcounter{footnote}{0}

%\baselineskip 6mm

%%%%%%%%%%%%%%%%%%%
%---------------SECTION-------------------%

\section{Introduction}

The standard model has a source of CP violation from the QCD dynamics~\cite{'tHooft:1976up}, which has however not been observed~\cite{Baker:2006ts}. This is so-called the strong CP problem.
One of the most attractive solutions to it is provided by the Peccei-Quinn (PQ) mechanism~\cite{Peccei:1977hh,Peccei:1977ur}, where anomalous $U(1)_\text{PQ}$ symmetry is introduced. The mechanism predicts the existence of a light pseudo-scalar field called axion~\cite{Weinberg:1977ma,Wilczek:1977pj}, which is also a candidate of dark matter in the universe~\cite{Preskill:1982cy,Abbott:1982af,Dine:1982ah}.

The axion model is however not always cosmologically safe. The QCD dynamics explicitly break the $U(1)_\text{PQ}$ symmetry down to a discrete subgroup $Z_{N_\text{DW}}$. If the subgroup is non-trivial (i.e.~$N_\text{DW}>1$), there exist stable domain walls which eventually dominate the energy density of the universe~\cite{Zeldovich:1974uw,Sikivie:1982qv}. The domain problem can be avoided if the PQ symmetry is already broken during inflation, but this scenario requires a small inflation scale to suppress the isocurvature perturbation~\cite{Linde:1984ti,Linde:1985yf,Seckel:1985tj,Lyth:1989pb,Lyth:1991ub,Turner:1990uz,Linde:1991km} or a flat potential of the PQ symmetry breaking field so that the effective PQ symmetry breaking scale is large during inflation~\cite{Linde:1991km}.

The domain wall problem can be solved if the $Z_{N_\text{DW}}$ symmetry is also explicitly broken and domain walls are unstable~\cite{Sikivie:1982qv}. Actually the PQ ``symmetry" is not at all symmetry since it is explicitly broken by the quantum anomaly. There would be no wonder that the PQ symmetry is also explicitly broken by a small amount at the classical level, and there is no residual $Z_{N_\text{DW}}$ symmetry. For example, if the PQ symmetry is accidental symmetry as a result of other exact symmetry~\cite{Lazarides:1985bj,Casas:1987bw,Randall:1992ut,Barr:1992qq,Holman:1992us,Dine:1992vx,Dias:2002gg,Choi:2009jt,Carpenter:2009zs,Harigaya:2013vja,Harigaya:2015soa,Fukuda:2017ylt}, we expect explicit PQ symmetry breaking by higher dimensional operators, which may also break the $Z_{N_\text{DW}}$ symmetry.%
\footnote{ This seems to be difficult to achieve in models with one-step symmetry breaking, as the exact symmetry results in stable topological defects. Models with several symmetry breakings like the one in Ref.~\cite{Barr:1992qq} can have a desired property. Ref.~\cite{Ringwald:2015dsf} considers one-field models, but the symmetry imposed there is anomalous and needs further completion of the model.}

The unstable domain wall mainly decays into axions,
%The abundance of the axions produced in this way may much larger than the one produced from the mis-alignment mechanism.
which can explain the observed amount of dark matter~\cite{Hiramatsu:2010yn,Hiramatsu:2012sc,Kawasaki:2014sqa} for a range of the decay constant $f_a$ much smaller than the one required for the misalignment mechanism~\cite{Preskill:1982cy,Abbott:1982af,Dine:1982ah} ($f_a > 10^{11}~\text{GeV}$). Such a small decay constant is relavent for several future searches for
solar axions~\cite{Vogel:2013bta,Armengaud:2014gea,Anastassopoulos:2017kag} and halo axions~\cite{Rybka:2014cya,Hochberg:2016ajh,TheMADMAXWorkingGroup:2016hpc,Arvanitaki:2017nhi}.
For a recent review of axion searches see e.g.~\cite{Irastorza:2018dyq}.
This scheme is however strongly constrained by non-observation of the effect of the strong CP phase. The explicit PQ symmetry breaking gives a small axion mass in addition to the one given by the QCD strong dynamics, and hence the cancellation of the strong CP phase becomes incomplete.
The large enough explicit PQ symmetry breaking to obtain the dark matter abundance tends to yield too much strong CP phase unless $f_a$ is close to the astrophysical lower bound, or the phase of the explicit symmetry breaking term is accidentally small.
%In other word, if the magnitude of the explicit breaking is small enough to suppress the strong CP phase, the domain walls are too long-lived and produce too much axions.

In the previous works~\cite{Hiramatsu:2010yn,Hiramatsu:2012sc,Kawasaki:2014sqa} it is assumed that the universe is radiation-dominated after the QCD phase transition.
In this letter we investigate the case where the universe is matter-dominated around the temperature of the MeV scale and domain walls decay during this matter dominated epoch. Such a cosmological scenario is actually expected if there is a very weakly coupled field such as a moduli field.
The matter dominance results in the dilution of the axions emitted from domain walls,
% relative to the usual case of the radiation dominance,
and hence the required magnitude of the explicit PQ symmetry breaking becomes smaller. We show how the viable parameter space is extended.

\section{Axions from long-lived domain walls}

We assume that the PQ symmetry is restored at some time after inflation and is spontaneously broken later. This include the cases where the Hubble scale during inflation is larger than the PQ symmetry breaking scale and thermal/non-thermal effects restore the PQ symmetry. Once the $U(1)_\text{PQ}$ symmetry is broken, cosmic strings are produced~\cite{Kibble:1976sj}. The reconnection and the decay of strings maintain the number of strings per horizon size roughly one, which is so-called the scaling law~\cite{Kibble:1984hp,Bennett:1985qt}.

As the temperature of the universe becomes as low as the QCD phase transition temperature, the explicit breaking of the PQ symmetry by the QCD dynamics becomes effective, and domain-walls are formed between the strings, with the wall-tension given by~\cite{Hiramatsu:2012sc,Huang:1985tt}
\begin{align}
\sigma_\text{dw}\simeq 9 m_a f_a^2,
\end{align}
where $m_a$ is the axion mass. In the parameter space of interest the domain walls decay after the QCD phase transition. Thus we consider the axion mass at the zero-temperature
given by~\cite{Weinberg:1977ma}
\begin{align}
m_a \simeq 6~\text{meV} \frac{10^9~\text{GeV}}{f_a}.
\end{align}
If there remains a discrete subgroup $Z_{N_\text{DW}} (N_\text{DW} > 1)$ of the PQ symmetry, the domain wall-string network is stable. This is for example the case with the DFSZ model~\cite{Dine:1981rt,Zhitnitsky:1980tq}. The network roughly follows the scaling-law~\cite{Hiramatsu:2010yn,Press:1989yh}, with the energy density given by
\begin{align}
\rho_\text{dw} (t) \simeq {\cal A} \frac{\sigma_\text{dw}}{t}.
\end{align}
Here ${\cal A}$ is an $O(1)$ factor which can be determined by numerical simulations.
The energy density of the network decreases to the second power of the scale factor of the universe, and eventually dominates the universe. This is inconsistent with the success of the standard cosmology, and is called the domain wall problem~\cite{Sikivie:1982qv}.

The problem may be solved if the $Z_{N_\text{DW}}$ symmetry is explicitly broken so that the domain walls are unstable.
%This is in fact expected from a theoretical point of view on the PQ symmetry. A notable feature the PQ mechanism is that the PQ symmetry is NOT a symmetry against the QCD interaction, while it must be an extremely good symmetry in order to suppress the strong CP phase. Such a weird``symmetry" may be well understood if the PQ symmetry is an accidental symmetry as a result of other exact symmetry~\cite{xxx}. This is analogous to the baryon symmetry of the standard model which accidentally holds because of the $SU(3)_c$ gauge symmetry, while is explicitly broken by the anomaly against the electroweak gauge interaction.
%See~\cite{xxx} for models of the PQ mechanism where the PQ symmetry is just an accidental one. In such a setup the origin of the stability of domain walls, the discrete subgroup $Z_{N_\text{DW}}$, may be explicitly broken by a small amount e.g.~by higher dimensional operators.
We parametrize the effect of the explicit breaking by the following bias term
\begin{align}
V= - 2\Xi v^4 \text{cos}\left( \frac{a}{v} + \delta \right),
\end{align}
where $v = N_\text{DW} f_a$ is the PQ symmetry breaking scale, $\delta$ and $\Xi$ are constants. We use the phase convention where the QCD non-perturbative effect creates the minima at $a = 2\pi k \times f_a  (k= 0,1\mathchar`-N_\text{DW}-1)$.
This bias term resolves the degeneracy between the $N_\text{DW}$ minima, and puts pressure on domain walls. The network collapses around the time $t_\text{dec}$ when the pressure beats the tension of domain walls~\cite{Kawasaki:2014sqa},
\begin{align}
t_\text{dec} \simeq C_d \frac{{\cal A} \sigma_\text{dw}}{\Xi v^4 \left( 1-\text{cos}\left( 2\pi / N_\text{DW}\right) \right)  },
\end{align}
where $C_d$ is a numerical constant.

The energy of the domain walls is transferred into axions. The resultant number density of the axion is given by
\begin{align}
\label{eq:na}
n_a (t_\text{dec}) \simeq \frac{\rho_\text{dw} (t_\text{dec})}{\tilde{\epsilon}_a m_a},
\end{align}
where $\tilde{\epsilon}_a$ is the average energy of the radiated axions normalized by the axion mass.
We find that Eq.~(\ref{eq:na}) reproduces the number density estimated by numerically solving the equation of motion of the domain wall energy density and the axion number density with an accuracy of few ten percents, if we determine $t_\text{dec}$ (i.e.~$C_d$) by the ``10\% criteria" in Ref.~\cite{Kawasaki:2014sqa}.

In order to suppress the axion abundance, the bias term must be sufficiently large so that the domain walls decay early enough.
The bias term, which explicitly breaks the PQ symmetry, results in a non-zero strong CP phase. In the limit where the bias term is much smaller than the potential given by the QCD non-perturbative effect, the phase is given by
\begin{align}
\theta \simeq  - 2 N_\text{DW}^3 \Xi \text{sin}\delta \frac{f_a^2}{m_a^2}.
\end{align}
The relation between the strong CP phase and the neutron electric dipole moment is estimated by various methods and approximation~\cite{Baluni:1978rf,Crewther:1979pi,Kanaya:1981se,Cea:1984qv,Schnitzer:1983pb,Musakhanov:1984qy,Hong:2007tf}, and the estimation are different from each others by a factor of $O(10)$.
This results in the uncertainly in the upper bound on $\theta$. We conservatively accept the bound of $|\theta| < 10^{-10}$.

\subsection{Axion abundance without entropy production}

Now we evaluate the axion abundance and the required magnitude of $\Xi$, assuming that the universe is radiation-dominated below the QCD phase transition temperature.
%We use the approximation that the domain walls suddenly decay at $t = t_\text{dec}$. As the decay is actually rapid~\cite{Kawasaki:2014sqa}, the expected error due to this approximation is few ten percents.
The axion abundance is estimated as~\cite{Kawasaki:2014sqa}
\begin{align}
\frac{\rho_a}{s} \simeq 7\times 10^{-10} \text{GeV} \left( \frac{10^9~\text{GeV}}{f_a} \right)^{1/2} \left( \frac{10^{-50}}{\Xi} \right)^{1/2}  \frac{{\cal A}^{3/2} C_d^{1/2}}{\tilde{\epsilon}_a N_\text{DW}^2 \text{sin} (\pi / N_\text{DW})},
\end{align}
where $s$ is the entropy density.
The observed dark matter abundance, $\rho_\text{DM} /s \simeq 4 \times 10^{-10}$ GeV, is reproduced with the magnitude of the explicit PQ symmetry breaking given by
\begin{align}
\label{eq:bias_RD}
\Xi \simeq 2\times 10^{-50}  \frac{10^9~\text{GeV}}{f_a} \frac{{\cal A}^3 C_d}{\tilde{\epsilon}_a^2  N_\text{DW}^4 \text{sin}^2 (\pi / N_\text{DW})},
\end{align}
which yields a non-zero strong CP phase,
\begin{align}
\theta \simeq 9 \times 10^{-10}  \text{sin}\delta  \left( \frac{f_a}{10^9~\text{GeV}} \right)^{3} \frac{{\cal A}^3 C_d}{\tilde{\epsilon}_a^2  N_\text{DW} \text{sin}^2 (\pi / N_\text{DW})}.
\end{align}
Using Eq.~(\ref{eq:bias_RD}), the temperature at which the domain walls decay is given by
\begin{align}
\label{eq:temp_DWdec}
T_\text{dec,DW} \simeq 30~\text{MeV} \frac{f_a}{10^9~\text{GeV}} \frac{{\cal A}}{\tilde{\epsilon}_a}.
\end{align}

\begin{figure}[htbp]
\centering
\includegraphics[width=0.6\linewidth]{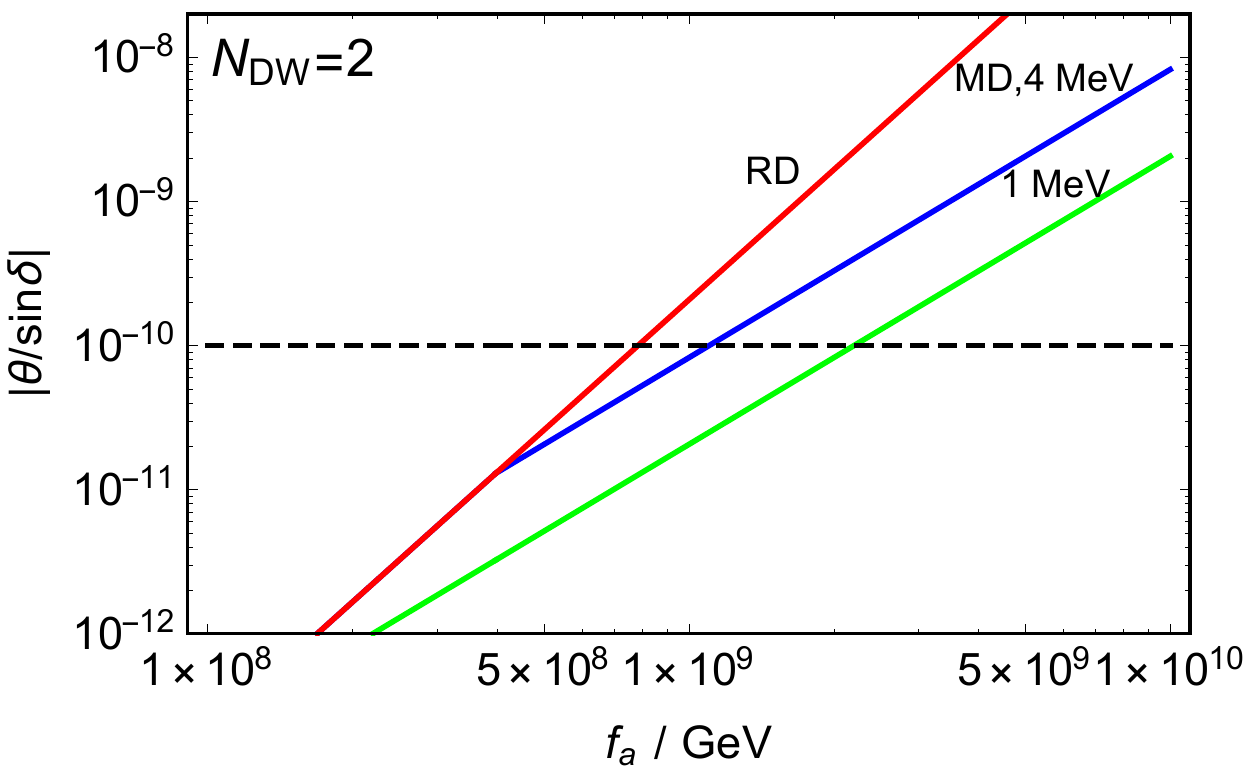}
\includegraphics[width=0.6\linewidth]{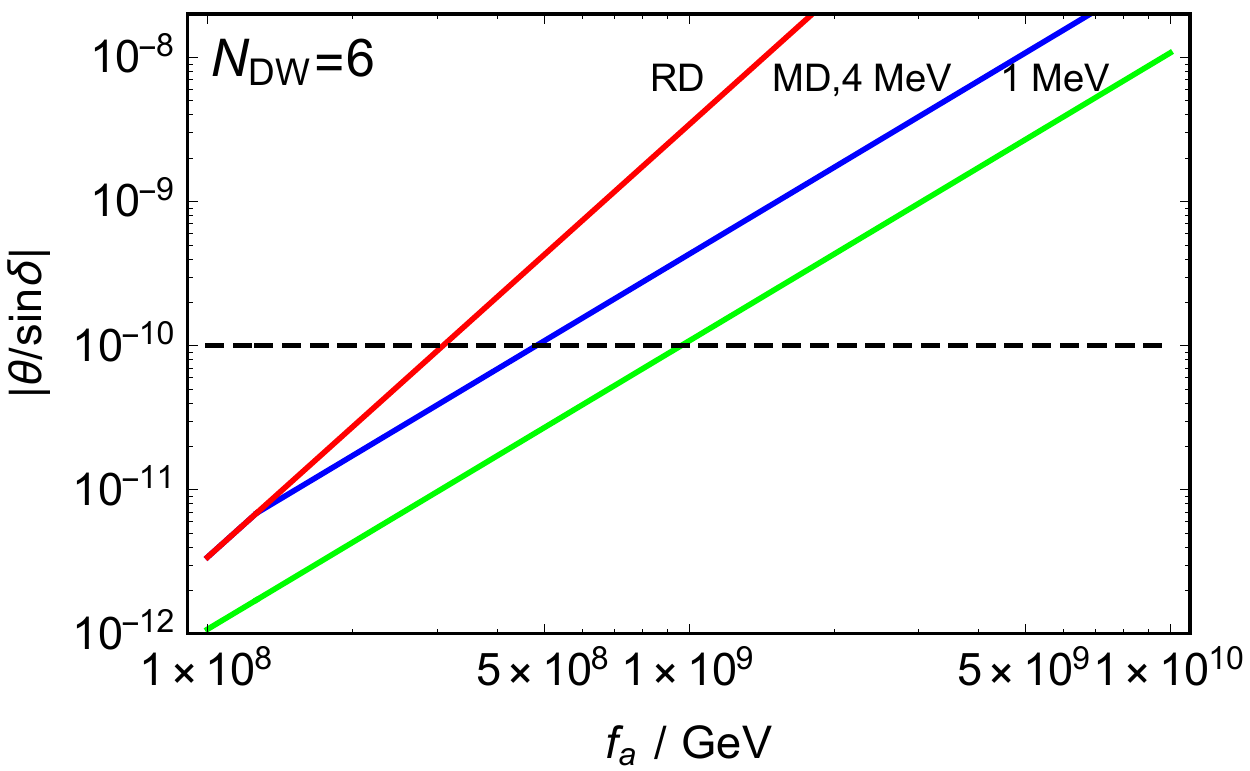}
\caption{\textsl{\small
The prediction on the strong CP phase as a function of $f_a$. The red lines assume that the universe is radiation-dominated after the QCD phase transition. The blue (green) lines assume that the universe is matter-dominated when the domain wall decays, and the matter decays at the temperature of 4 (1) MeV.
}}
\label{fig:phase}
\end{figure}

In Fig.~\ref{fig:phase}, we show the prediction on the strong CP phase by a red solid line.
We list the numerical values of ${\cal A}$, $C_d$ and $\tilde{\epsilon}_a$ taken from~\cite{Kawasaki:2014sqa} in Table~\ref{tab:constants}.
For $N_\text{DW}=2$, to satisfy the upper bound $|\theta|< 10^{-10}$, $f_a < 8 \times 10^8 $ GeV is required for ${\sin} \delta \sim 1$.
Such parameter region is disfavored by the cooling of white dwarfs~\cite{Raffelt:2006cw}, if there exist a significant axion-electron coupling as is the case with the DFSZ model.
%That parameter region seems to be constrained by the cooling of the SN1987~\cite{xxx}. Note however that the constraint involves astrophysical uncertainties e.g.~ the modeling of the structure of the SN.
Accidentally small $\delta$ allows for larger $f_a$. For example, if $\delta < 0.1$, $f_a < 2 \times 10^9 $ GeV is allowed.%
\footnote{Ref.~\cite{Kawasaki:2014sqa} assumes an agressive bound $|\theta| < 7\times 10^{-12}$ and finds much more restricted results.}
The constraint is stronger for $N_\text{DW}=6$. This is because of a larger ${\cal A}$ for larger $N_\text{DW}$ and hence larger domain wall energy, which can understood by the fact that more domain walls are attached to a string. To satisfy the bound on the strong CP phase, $f_a < 3 \times 10^8 $ GeV is required for ${\sin} \delta \sim 1$.
%, while the matter-dominance allows for $f_a < 10^9 $ GeV.

%%%
\begin{table}[tb]
\caption{Numerical values taken from~\cite{Kawasaki:2014sqa}.}
\begin{center}
\begin{tabular}{|c|c|c|c|}
\hline
 & ${\cal A}$ & $C_d$ & $\tilde{\epsilon}_a$ \\ \hline
 $N_\text{DW}=2$ & 0.7 & 5.3 & 2.0   \\
  $N_\text{DW}=6$ & 2.2 & 2.1 & 2.0   \\ \hline
\end{tabular}
\end{center}
\label{tab:constants}
\end{table}%

%%%

%%%%%%%%%%%%%%%%%%%%%%%%%%%%%%%%%%%%%%%%%%%%%%%%%%%%%%%%%%%%%
%%%%%%%%%%%%%%%%%%%%%%%%%%%%%%%%%%%%%%%%%%%%%%%%%%%%%%%%%%%%%%%%%%%%%%%%

\subsection{Axion abundance with entropy production}
As is shown in Eq.~(\ref{eq:temp_DWdec}), for the parameters which reproduce the observed dark matter abundance,
% from the axions produced from domain walls,
%for a radiation-dominated universe,
the domain walls decay above the temperature of the MeV scale.
The era is before the onset of the Big-Bang Nucleosythesis, and the universe may be matter-dominated.
%, namely the temperature of the MeV scale.
%In this case the domain walls may decay during the matter-dominated era.
Then the axion abundance from the domain walls is smaller than the simple case of the radiation dominance and the predicted strong CP phase is suppressed.

Actually if there exists a light, very weakly interacting particle, after the inflaton decays and the universe becomes radiation-dominated, the particle eventually dominates the energy density of the universe.
%We assume a field which dominates the energy density of the universe and decays around the MeV scale.
We call such a particle as a moduli, although it is not necessarily a scalar particle.
For example, a perticle with a mass $m$ which couples to the standard model particle with a dimension-5 operator suppressed by a scale $M_*$ decays around the temperature
\begin{align}
T_{\text{dec},m} \sim 10 \text{MeV} \left( \frac{m}{10^5~\text{GeV}} \right)^{3/2} \frac{10^{18}~\text{GeV}}{M_*}.
\end{align}
A gravitationally coupled particle ($M_*\sim 10^{18}$ GeV) with a mass around 100 TeV gives $T_{\text{dec},m}\sim$ MeV.
In the supersymmetric standard model with the supersymmetry breaking mediated by gravitational interaction, the masses of moduli fields are as large as the masses of the scalar partners of the standard model fermions. The scalar masses of $O(100)$ TeV can explain the observed higgs mass of 125 GeV~\cite{Okada:1990vk,Ellis:1990nz,Haber:1990aw,Hall:2011jd,Ibe:2011aa}.
If $M_* \sim f_a$, which is the case for the radial direction of the PQ symmetry breaking field, $m=O(0.1\mathchar`-1)$ GeV yields a MeV scale $T_{\text{dec},m}$.

When the domain walls decay, the ratio of the number density of the axion to the energy density of the moduli, $\rho_m$, is given by
\begin{align}
\frac{n_a}{\rho_m} = \frac{3 n_a(t_\text{dec})}{ 4 \mpl / t_\text{dec}^2 },
\end{align}
which does not change under the expansion of the universe until the moduli decays. Converting $\rho_m$ to the entropy density when the moduli decays (see the appendix), we obtain
\begin{align}
\label{eq:abundance_md}
\frac{\rho_a}{s} \simeq 1.2 T_{\text{dec},m} m_a\frac{n_a}{\rho_m} =  4\times 10^{-11} \text{GeV} \left( \frac{10^9~\text{GeV}}{f_a} \right)^{2}  \frac{10^{-50}}{\Xi} \frac{ T_{\text{dec},m} }{1~\text{MeV}}   \frac{{\cal A}^2 C_d}{\tilde{\epsilon}_a N_\text{DW}^4 \text{sin}^2 (\pi / N_\text{DW})},
\end{align}
where $T_{\text{dec},m}$ is defined by the decay rate of the moduli $\Gamma$ as
\begin{align}
\label{eq:Tdec_def}
\Gamma = 3 H(T_{\text{dec},m}),
\end{align}
with the Hubble scale evaluated by that during radiation dominated era.
%See the appendix for the derivation of the first equality in Eq.~(\ref{eq:abundance_md}).
%%
The dark matter abundance is obtained for
\begin{align}
\Xi \simeq 1\times 10^{-51}  \left( \frac{10^9~\text{GeV}}{f_a} \right)^2 \frac{ T_{\text{dec},m} }{1~\text{MeV}}  \frac{ {\cal A}^2 C_d }{ \tilde{\epsilon}_a  N_\text{DW}^4 \text{sin}^2 (\pi / N_\text{DW})},
\end{align}
which yields a non-zero strong CP phase,
\begin{align}
\theta \simeq 3 \times 10^{-11}  \text{sin}\delta  \left( \frac{f_a}{10^9~\text{GeV}} \right)^{2}  \frac{ T_{\text{dec},m} }{1~\text{MeV}}  \frac{{\cal A}^2 C_d}{\tilde{\epsilon}_a  N_\text{DW} \text{sin}^2 (\pi / N_\text{DW})}.
\end{align}

The temperature $T_{\text{dec},m}$ is constrained to be $T_{\text{dec},m}> 0.7$ MeV, as the low temperature leads to incomplete thermalization of neutrinos, affecting the Big-Bang nucleosynthesis~\cite{Kawasaki:2000en}\footnote{%%
If the moduli decays into hadrons with significant hadronic branch the BBN gives more stringent constraint as $T_{\text{dec},m} > 3\text{--}4$~MeV~\cite{Kawasaki:2000en}.}.
This also reduces the energy density of the relativistic component of the universe, which changes the expansion history of the universe and affects the spectrum of the cosmic microwave background. The Planck satellite obtains the constraint $N_\text{eff}> 2.8$ (95\%C.L.)~\cite{Ade:2015xua}. Comparing this with the results in~\cite{Kawasaki:2000en}, we obtain $T_{\text{dec},m}> 4$ MeV. This bound can be evaded if the moduli has a significant branching fraction to radiations such as neutrinos or axions. We consider two reference values $T_{\text{dec},m}=1$ and $4$ MeV.

The prediction on the strong CP phase is shown in Fig.~\ref{fig:phase} by the solid blue and green lines, with $T_\text{dec,m} = 4$ and 1 MeV, respectively. For $N_\text{DW} =2$ with $T_{\text{dec},m}=1$ MeV, the fine-tuning of $\delta$ is not necessary as long as $f_a< 2\times 10^9$ GeV. For that large $f_a$, fine-tuning of about 10 \% is required without the matter domination. Allowing 10 \% tuning, $f_a$ may be as large as $7\times 10^{9}$ GeV.
%, which requires 1 \% tuning without the matter domination.

\section{Summary and discussion}
We comment on the production mechanisms of axion dark matter for a small decay constant discussed in the literature. If the misalignment angle is very close to the maximal one, the un-harmonic effect delays the commencement of the oscillation of the axion. This enhances the axion abundance and reproduces the observed dark matter abundance even if $f_a \ll 10^{11}$ GeV~\cite{Lyth:1991ub,Turner:1985si,Visinelli:2009zm}. This scenario requires the fine-tuning of the misalignment angle, as well as a very small inflation scale to suppress the isocurvature perturbation.
If the radial direction of the PQ symmetry breaking field takes a large value in the early universe, the production of the axions by the parametric resonance~\cite{Kofman:1994rk,Kofman:1997yn} can produce axion dark matter~\cite{Co:2017mop}. The scenario requires that the potential of the PQ symmetry breaking field is flat, which would call for supersymmetry.
Kinetic energy domination around the QCD phase transition enhances the axion abundance produced by the misalignment mechanism and allows for a small decay constant~\cite{Visinelli:2009kt}.

In this paper we have investigated the production of axions from unstable domain walls because of the explicit PQ symmetry breaking, assuming that
the universe is matter-dominated around the temperature of the MeV scale and domain walls decay during this matter dominated epoch.
This scenario can explain the observed amount of dark matter by the axion
%even if the decay constant is around $10^9$ GeV,
without tuning the phase of the explicit breaking as long as $f_a < O(10^{9})$ GeV. Allowing the tuning of $O(10)$\%, $f_a < O(10^{10})$ GeV is allowed.

%We consider a very low temperature of $O(1)$ MeV at which the matter field decays. This can be naturally explained from a gravitationally coupled field with a mass around 100 TeV, of the radial direction of the PQ symmetry breaking field with a mass $O(0.1\mathchar`-1)$ GeV.

\section*{Acknowledgements}

This work was supported by
the Director, Office of Science, Office of High Energy and Nuclear Physics, of the US Department of Energy under Contract DE-AC02- 05CH11231 (K.H.), the National Science Foundation under grants PHY-1316783 and PHY-1521446 (K.H),
JSPS KAKENHI Grant Nos.~17H01131 (M.K.) and 17K05434 (M.K.), MEXT KAKENHI Grant No.~15H05889 (M.K.), and World Premier International Research Center Initiative (WPI Initiative), MEXT, Japan (M.K.).

\appendix

\section{Derivation of Eq.~(\ref{eq:abundance_md})}

The equations governing the evolution of the energy densities of the moduli, $\rho_m$, and radiation, $\rho_r$, and the scalar factor, $a$, is given by
\begin{align}
\dot{\rho_m} + 3 H \rho_m = - \Gamma \rho_m,\\
\dot{\rho_r} + 4 H \rho_r = + \Gamma \rho_m, \\
\frac{\dot{a}}{a} = H = \frac{1}{\sqrt{3} \mpl} \sqrt{\rho_m + \rho_r},
\end{align}
with the initial conditions given at $t_i \ll 1/ \Gamma$,
\begin{align}
\rho_m(t= t_i) \equiv \rho_{m,i}, \\
\rho_r(t= t_i) = 0, \\
a(t= t_i) \equiv a_i
\end{align}
After the change of variables
\begin{align}
t = x / \Gamma,~\rho_m = a^{-3} m \Gamma^2 \mpl^2,~\rho_r = a^{-4} r \Gamma^2 \mpl^2,
\end{align}
the equations are given by
\begin{align}
\frac{\text{d} m}{\text{d} x} = -  m,\\
\frac{\text{d} r}{\text{d} x} = a m, \\
\frac{\text{d} a}{\text{d} x} = \frac{a}{\sqrt{3}} \sqrt{a^{-3} m + a^{-4} r}.
\end{align}

The solution for $m$ is
\begin{align}
m = e^{-(x-x_i)} m_i,
\end{align}
where $m_i \equiv m(x_i)$. The evolution equations of $r$ and $a$ are given by
\begin{align}
\label{eq:drdx}
\frac{\text{d} r}{\text{d} x} = a e^{-(x-x_i)} m_i,\\
\label{eq:dadx}
\frac{\text{d} a}{\text{d} x} = \frac{a}{\sqrt{3}} \sqrt{a^{-3} e^{-(x-x_i)} m_i + a^{-4} r}.
\end{align}
For $x \rightarrow \infty $, $r$ becomes a constant $\equiv r_f$.

The quantity we are interested in is
\begin{align}
\frac{n_a}{s} (t \rightarrow \infty) = \frac{n_a}{\rho_r^{3/4}} (t \rightarrow \infty)  \frac{\rho_r^{3/4}}{s} = \frac{n_a}{\rho_m} (t= t_i) \frac{m_i}{r_f^{3/4}} \sqrt{\Gamma \mpl}  \frac{\rho_r^{3/4}}{s}.
\end{align}
Using the definition of $T_\text{dec,m}$ in Eq.~(\ref{eq:Tdec_def}), we obtain
\begin{align}
\frac{n_a}{s} (t \rightarrow \infty) =  \frac{n_a}{\rho_m} (t= t_i) T_\text{dec,m} \frac{m_i}{r_f^{3/4}} \frac{3^{5/4}}{4}.
\end{align}
A numerical evaluation of Eqs.~(\ref{eq:drdx}) and (\ref{eq:dadx}) gives $m_i/r_f^{3/4} \simeq 1.2$, and we obtain
\begin{align}
\frac{n_a}{s} (t \rightarrow \infty) \simeq 1.2 \times T_\text{dec,m}  \frac{n_a}{\rho_m} (t= t_i).
\end{align}

\end{document}